\documentclass{article}

\usepackage[english]{babel}
\usepackage{amsmath,amssymb}

\usepackage[letterpaper,top=2cm,bottom=2cm,left=3cm,right=3cm,marginparwidth=1.75cm]{geometry}

\usepackage{amsmath}
\usepackage{graphicx}
\usepackage[colorlinks=true, allcolors=blue]{hyperref}

\title{Highly-Multimode Solitons in Step-Index Optical
Fiber}
\author{Yuhang Wu, Nicholas Bender, Demetrios N. Christodoulides, and Frank W. Wise}

\begin{document}
\date{}
\maketitle

\begin{abstract}
  We report the generation of multimode solitons in step-index fiber. The solitons are superpositions of 5-10 temporally-aligned transverse modes, they exhibit speckled beam profiles, and the spatio-spectral variation across the soliton can be complex. Greater understanding of multimode solitons should create a foundation for further research into complex multimode nonlinear phenomena in step-index fibers. With energies of tens of nanojoules and durations around 100 fs, the multimode solitons are speckled fields with peak powers that can exceeed 300 kW, parameters which may be valuable in applications. 
\end{abstract}

A soliton is a wave packet that maintains its shape owing to the balance of nonlinear and linear dispersive or diffractive effects \cite{Agrawal2000}. Single-mode fiber (SMF) is an ideal platform for studying solitons governed by the Nonlinear Schr\"{o}dinger Equation (NLSE) \cite{Hasegawa1973,Mollenauer1980}. The study of solitons in SMF has, in return, greatly facilitated the understanding of a wide variety of phenomena including supercontinuum generation \cite{Herrmann2002,Skryabin2010} mode-locked lasers \cite{Mollenauer1984,Haus1985}, and optical communication \cite{Hasegawa2000}. In multimode (MM) fibers, pulse propagation involves both spatial and temporal degrees of freedom \cite{Mafi2012}, enabling more complex solitons~\cite{Hasegawa1980}. 

Theoretical predictions of solitons in MM fiber date back to the 1980s \cite{Hasegawa1980,Crosignani1981,Yu1995,Chien1996,Raghavan2000,Lederer2008}. Grudinin et al. inferred soliton formation from early experiments in MM graded-index (GRIN) fiber. An input pulse underwent fission along with stimulated Raman scattering to yield a Stokes-shifted soliton in the fundamental mode of the fiber \cite{Grudinin1988}. Systematic studies of solitons in MM fiber have only been undertaken since 2013 \cite{Zitelli2021,Wright2015,Sun2022}. Renninger and Wise reported theoretical and experimental observations of solitons with energies in the lowest few modes of GRIN MM fiber and their Raman shifting \cite{Renninger2013}. Later, Wright et al. and Zitelli et al. provided more detailed studies for the generation of MM solitons  \cite{ Wright2015, Zitelli2021}. Zitelli et al. proposed the walk-off soliton theory and showed that the generated multimode soliton pulse duration and energy only depend on the coupling conditions and linear dispersive properties of the fiber \cite{Zitelli2021}. All the studies mentioned above are based on GRIN MM fibers.

Step-index MM fibers are the most common type of MM fibers, which motivates investigation of MM solitons in them. However, the large modal dispersion typical of step-index fibers presents a major challenge to the generation of MM solitons. Single-mode solitons, which can be understood with the ordinary nonlinear Schr\"{o}dinger equation, have been generated in MM step-index fiber \cite{rishoj2019}. Zitelli et al. have demonstrated the formation of solitons involving a few low-order modes in step-index fiber, and the energy eventually transferred to the fundamental mode \cite{Zitelli2022}. It is still an open question whether highly-multimode solitons can be generated in step-index fibers, and if so, what conditions lead to their formation. Some properties of step-index fibers should be conducive to the experimental investigation of MM solitons. The transverse modes of step-index fibers overlap more than the modes of GRIN fiber, which will facilitate the observation of speckled spatial profiles that would be the signature of a superposition of higher-order modes.  In addition, Raman beam cleanup \cite{Terry2007}, which hinders observation of MM solitons in GRIN fiber, does not occur in step-index fiber. Understanding of MM solitons in step-index fiber can be expected to underpin understanding of complex nonlinear phenomena such as supercontinuum generation and spatiotemporal mode-locking in MM fiber lasers \cite{Wu2022}.

Here we report theoretical and experimental studies of highly-multimode solitons in step-index fiber. We exploit fission and Raman scattering of high-power pulses to generate MM solitons in step-index fiber with modest modal dispersion. Numerical simulations agree qualitatively with the experimental results and help reveal that the observed solitons are spatio-temporally simple but can be spatio-spectrally complex. Solitons with peak power as high as 370 kW are generated. 

We simulate MM soliton formation and propagation by solving the generalized multimode nonlinear Schr\"{o}dinger equation~\cite{Poletti2008} using a parallel mode-based algorithm~\cite{Wright2017}. Specifically, we model the scalar-wave propagation through 10 meters of step index fiber, with a core diameter of 105 \textmu m (numerical aperture NA = 0.1). This fiber supports 105 modes per polarization (210 total) for wavelengths around 1550 nm. The modal delay with respect to the fundamental mode increases linearly with mode number, with an average difference between adjacently-indexed modes of 103 fs/m (see supplementary materials). These parameters are chosen to mirror the fiber used in the experiments described below. A Gaussian pulse at 1550 nm with full-width at half-maximum (FWHM) duration of 500 fs is launched into the fiber. The pulse energy (200 nJ) is evenly distributed across modes 16 through 25, with a randomly assigned phase for each mode. For statistically-equivalent launching conditions, qualitatively similar results are observed. Under these conditions, the dispersion length is $\sim 3$ m and the nonlinear length is $\sim 9$ cm when calculated using the fundamental mode. To keep the computation time reasonable, only the first 30 modes are included in the simulation.

\begin{figure}[hthb]
\centering
\includegraphics[width=\linewidth]{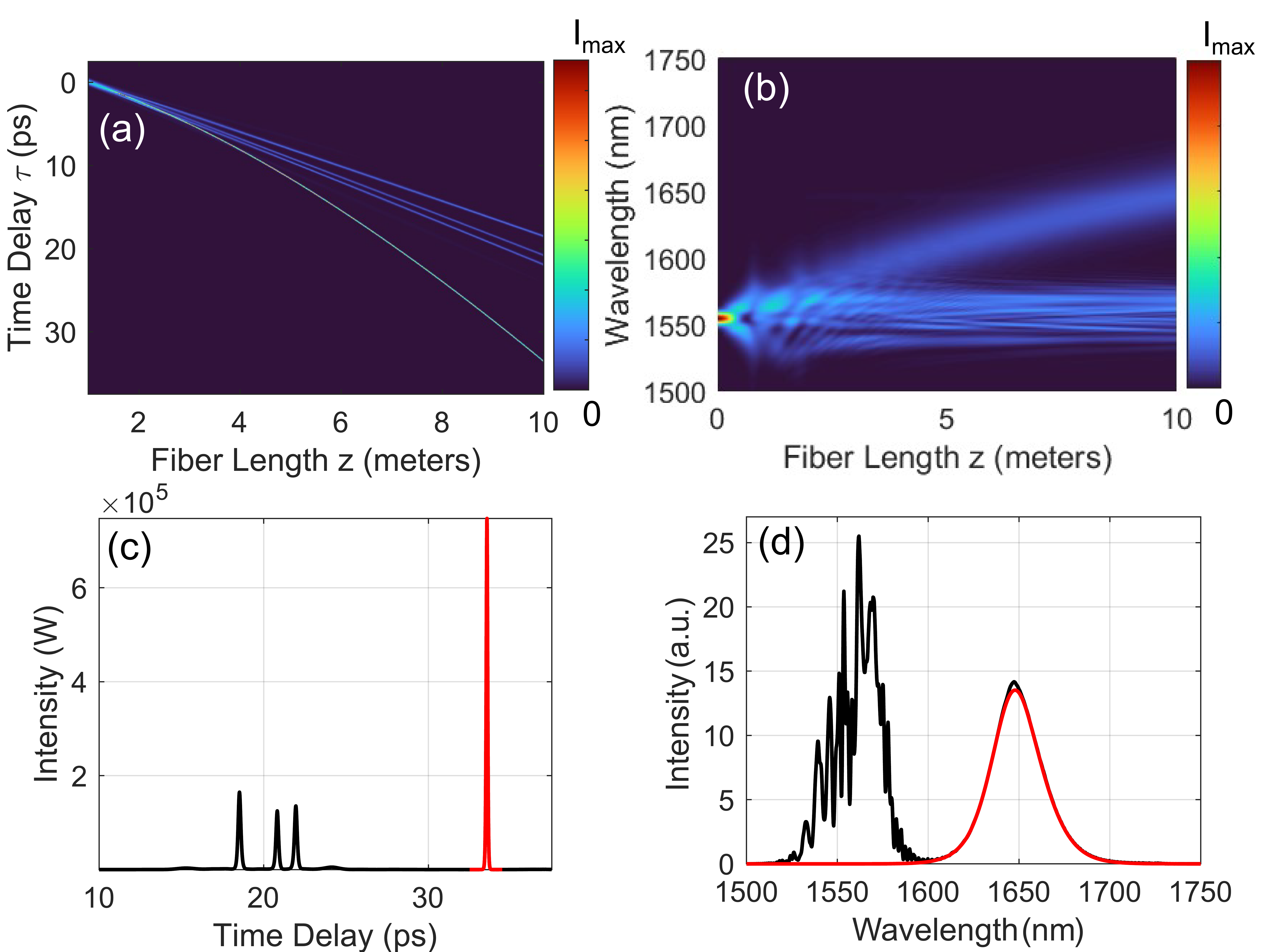}
\caption{Numerical simulation of MM soliton formation. (a) Spatially-integrated time-domain intensity evolution. (b) Spatially-integrated spectral-domain evolution. (c) Spatially-integrated power at the end of the 10-m fiber. (d) Spatially-integrated spectrum at the end of the fiber. The red plot corresponds to the spectrum of the MM soliton.}
\label{Figure_1_Whole_TS}
\end{figure}

The spatially-averaged ($\hat{x}, \hat{y}$) intensity of the optical field is shown as a function of time delay with respect to the pulse peak at fundamental mode at 1550 nm ($\tau$) vs. propagation ($z$) in Fig. \ref{Figure_1_Whole_TS}(a), and as a function of wavelength ($\lambda$) vs. propagation in Fig. \ref{Figure_1_Whole_TS}(b). At the beginning of the fiber, (0-1 m), the pulse undergoes compression in time. Simultaneously, the spectral bandwidth rapidly broadens to  $\sim 60$ nm. With further propagation (1-10 m) the pulse undergoes temporal fission, which creates multiple MM solitons. These solitons have 100-fs duration, which corresponds to $\sim 12$ cm dispersion length. They maintain their temporal shape and duration over more than 70 dispersion lengths in the fiber. Concurrently --in the spectral domain-- stimulated Raman scattering creates a constantly-shifting peak. The continuously red-shifting Raman peak decreases the group velocities of the modes in the solitons, thus introducing a significant time delay to the solitons. The residual light at 1550 nm corresponds to the energy dispersed away and a few single-mode or few-mode solitons. This nonlinear evolution starkly contrasts with the linear pulse evolution, where the wave packet in each mode broadens and temporally ``walks off’’ (see supplementary materials) from other modes owing to modal dispersion. While the observed behavior is reminiscent of soliton self-frequency shifting (SSFS) in SMF, it is important to keep in mind that this is a multimode process. Specifically, intermodal cross-phase modulation produces frequency shifts that compensate modal dispersion as part of soliton formation. At the end of the fiber a MM Raman soliton with a duration about 100 fs and peak power as high as 700 kW is generated (Figs. \ref{Figure_1_Whole_TS}(c,d)). We will focus on the Raman soliton, because it can be easily isolated for experimentally study. 

\begin{figure}[hthb]
\centering
\includegraphics[width=\linewidth]{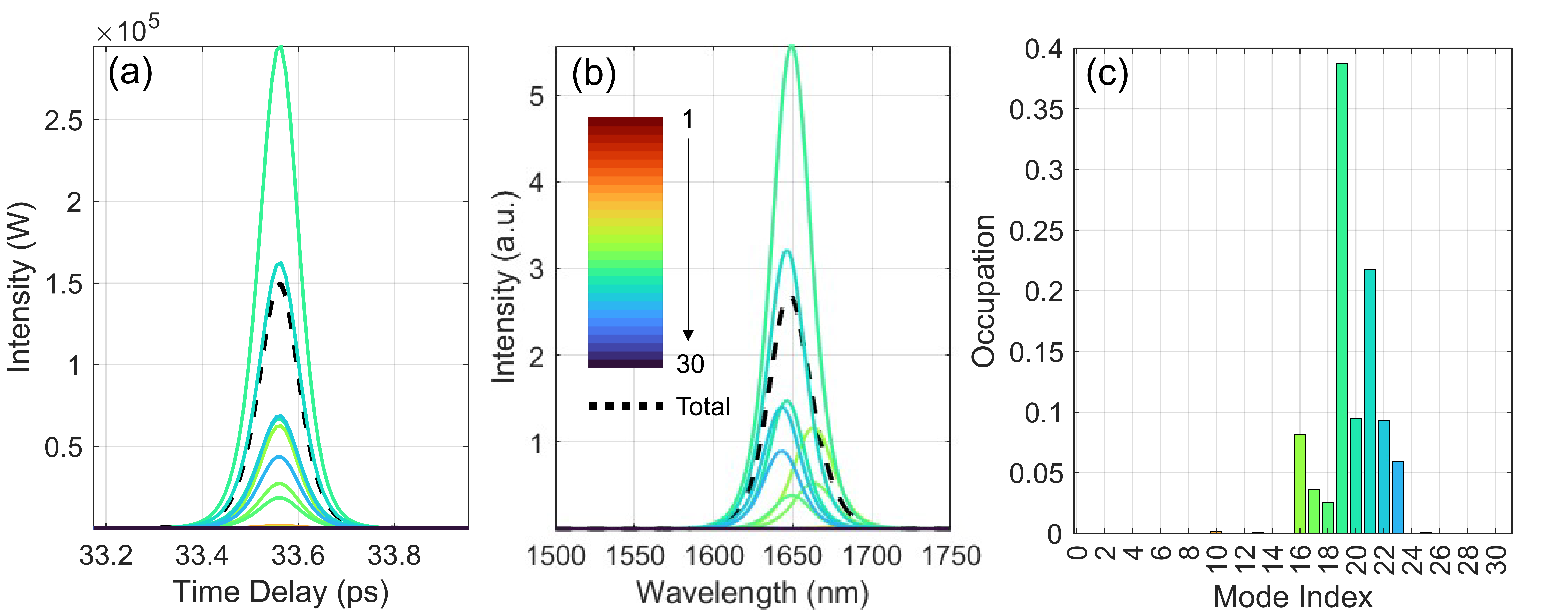}
\caption{Mode-resolved temporal (a) and spectral (b) profiles for the most red-shifted soliton in the numerical simulation of Fig. 1. Colors indicate modes according to the inset of (b). In (a), the dashed-black curve is the total intensity profile divided by a factor of 5. In (b), the dashed-black curve is the total spectral profile divided by a factor of 5.  (c) is the relative modal occupation.}
\label{Figure_2_Modal_TS}
\end{figure}

The mode-decomposed properties of the soliton with the largest time delay in Fig.~\ref{Figure_1_Whole_TS}(c) are shown in Fig.~\ref{Figure_2_Modal_TS}. In the temporal domain (Fig.~\ref{Figure_2_Modal_TS}(a)) the mode profiles have the same shape, the same time delay, and the same pulse width (100 fs). These properties are exhibited by the soliton when it forms and continue until the end of the fiber. In the spectral domain (Fig.~\ref{Figure_2_Modal_TS}(b)) the mode profiles are centered at different wavelengths, with up to 25 nm of separation. The mode-dependent shifts in the spectrum compensate for the large modal delays to keep the soliton together in time. The time-averaged modal populations (Fig.~\ref{Figure_2_Modal_TS}(c)) appear to be a random distribution of the launched modes. In contrast to previously-reported MM solitons 
 \cite{Wright2015,Zitelli2021}, the low-order modes (modes 1-15) have negligible populations. Raman beam cleaning causes power to transfer to the fundamental mode in GRIN fiber \cite{Terry2007}, which explains why lower-order modes eventually dominate MM solitons in GRIN fiber. The modal distribution of the Raman soliton observed here does not vary with propagation. 

\begin{figure}[hthb]
\centering
\includegraphics[width=\linewidth]{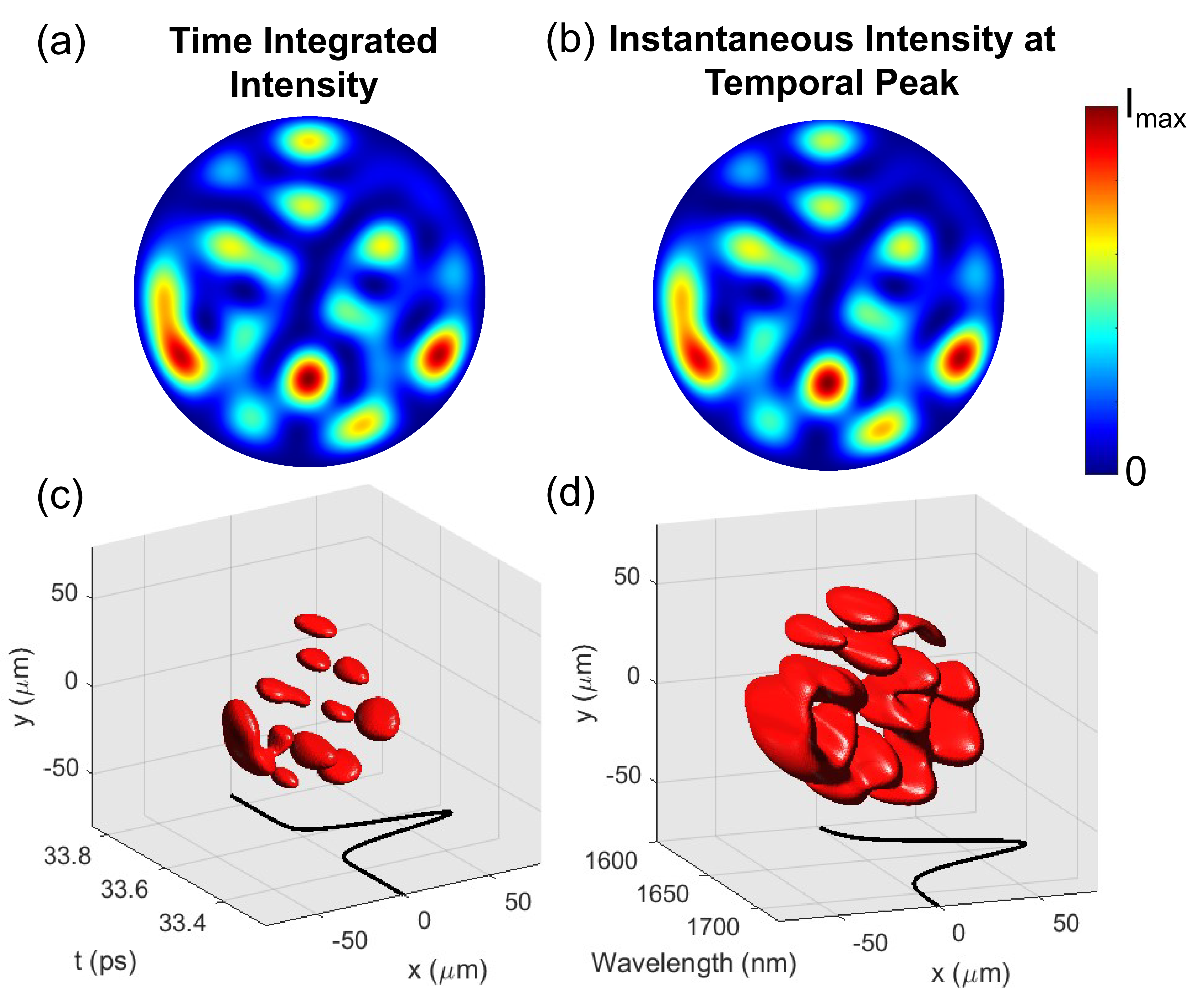}
\caption{Spatio-temporal and spatio-spectral profiles of MM soliton observed in numerical simulation. (a) is the time-integrated intensity pattern. (b) is the intensity pattern at the peak of the pulse. The edge of the profile is the core boundary. (c) is the spatio-temporal profile, represented by the isosurface with 10 per cent of the maximum intensity. (d) is the spatio-spectral profile, represented by the isosurface with 10 per cent of the maximum intensity.}
\label{Figure_3_Spatial_Beam}
\end{figure}

Fig.~\ref{Figure_3_Spatial_Beam} illustrates the spatial properties of the MM soliton presented in Figs. ~\ref{Figure_1_Whole_TS} and ~\ref{Figure_2_Modal_TS}. The primary feature is that the profile is speckled, as expected for a pseudo-random superposition of 8 high-order transverse modes. The time-integrated spatial intensity pattern at the output of the fiber (Fig.~\ref{Figure_3_Spatial_Beam}(a)) is almost the same as the spatial intensity pattern at the peak of the pulse (Fig.~\ref{Figure_3_Spatial_Beam}(b)). The speckled spatial profile of the soliton does not fluctuate significantly within the temporal envelope of the pulse. This is quantitatively reflected in the intensity contrast, defined as $C=\sqrt{\langle I(x,y)^2 \rangle / \langle I(x,y) \rangle ^ 2 - 1}$, where $I(x,y)$ is the spatial intensity pattern and $\langle \cdots \rangle$ indicates spatial averaging~\cite{Bromberg2014}. The intensity pattern at the peak of the pulse (Fig. ~\ref{Figure_3_Spatial_Beam}((b)) has higher contrast ($C=0.76$) than the time-integrated intensity pattern (Fig. ~\ref{Figure_3_Spatial_Beam}((a)) ($C=0.74$). Nevertheless, the two values are similar and both are much higher than the contrast of the time-integrated intensity of an equivalent linear propagation ($C=0.39$) (see supplementary materials). As such, the intensity contrast is useful for determining if a MM pulse propagating in SI is a soliton. For context, a uniform intensity pattern has $C=0$, while a monochromatic Rayleigh speckle pattern has $C=1$. The lack of temporal variation can be seen in Fig.~\ref{Figure_3_Spatial_Beam}(c), which shows the spatio-temporal profile of the pulse. The spectral variation of the pulse can be seen in the spatio-spectral profile of Fig.~\ref{Figure_3_Spatial_Beam}(d).

\begin{figure}[hthb]
\centering
\includegraphics[width=\linewidth]{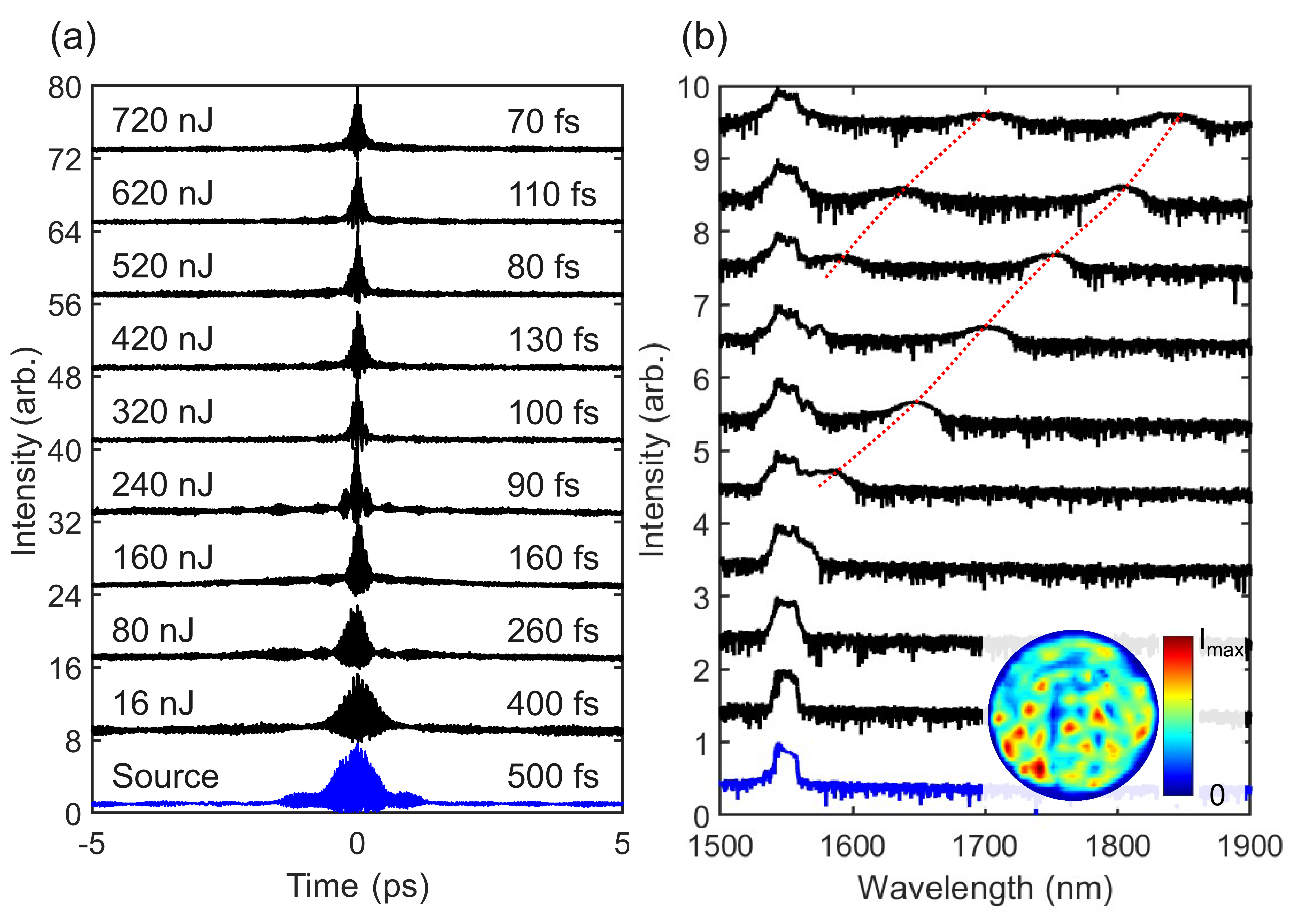}
\caption{Temporal and spectral measurements for different input pulse energy in the experiment. (a) Spatially integrated pulse measurements for fixed modal excitation. The input pulse energies are on the left of the figure, while the durations of the measured output pulses are on the right. The blue plot is the pulse directly from the laser source. (b) Spatially-integrated spectrum on the normalized logarithmic intensity scale. The red-dashed curves guide the eye to the Raman soliton peaks. The blue plot is the spectrum of the laser source. The inset is the output near-field profile at 16 nJ input pulse energy.}
\label{Figure_4_Exp_TSS}
\end{figure}

In the experimental setup, an erbium-doped fiber amplifier (Calmar model Cazadero) generates pulses with a temporal duration of $\sim 500$ fs FWHM, a spectral bandwidth of 8.3 nm FWHM, and a maximum pulse energy of 1 \textmu J (Fig. \ref{Figure_4_Exp_TSS}). The pulses are coupled into 10 meters of step-index MM fiber (Thorlabs FG105LVA) using a lens with focal length of 25 mm. Using a variable attenuator and a 3D stage to control the spot’s location on the fiber input face, we can generate a diverse range of input conditions. The fiber is loosely coiled, with a diameter of $\sim 35$ cm. The beam radius at the focus is 15 \textmu m. Since the fundamental mode radius is 40 \textmu m and the beam is launched with a deviation from the fiber center, high-order modes (HOMs) are excited. We observe that different excitation conditions always produce superpositions of HOMs. This is likely due to linear mode coupling caused by the clamps that hold the fiber in place and the coiling of the fiber. The output facet of the fiber is directly imaged by an InGaAs camera, via two lenses in a configuration with $11\times$ magnification. Using a flip-mirror, we measure the spatially-integrated temporal profile using an autocorrelator based on 2-photon photocurrent in a Si detector. The wavelength response of the detector covers all the wavelengths generated in the experiment. The beam is also coupled into a MM fiber and sent to an optical spectrum analyzer with a wavelength range of 1-2.5 \textmu m to record the spatially-integrated power spectrum.

Temporal and spectral characteristics of the output pulses for different input pulse energies are shown in Fig. \ref{Figure_4_Exp_TSS}. At the lowest energy (16 nJ), the pulse propagation is essentially linear, as indicated by the lack of spectral broadening. As the input pulse energy increases to 160 nJ, the spectrum broadens symmetrically and the autocorrelation narrows. At 160 nJ the spectrum becomes asymmetric owing to Raman scattering and the pulse autocorrelation narrows further. Given the emergence of Raman effects in the spectrum, soliton fission has likely occurred based on the numerical simulations. As the input pulse energy increases further, the pulse duration remains around 100 fs. For selected experiments, the autocorrelation was measured out to a delay of 20 ps, but a second peak was not observed (See supplementary materials). As is evident in Fig. \ref{Figure_4_Exp_TSS}(b), the first Raman peak in the spectrum continuously shifts to longer wavelengths, and a second Raman peak forms at an input energy of 520 nJ. We find that the output spatial beam pattern (for example, shown in the inset of Fig. \ref{Figure_4_Exp_TSS}(b)) and spectrum are very sensitive to the excitation conditions. Spatial intensity patterns observed for different input pulse energies are shown in the supplementary materials.

\begin{figure}[hthb]
\centering
\includegraphics[width=\linewidth]{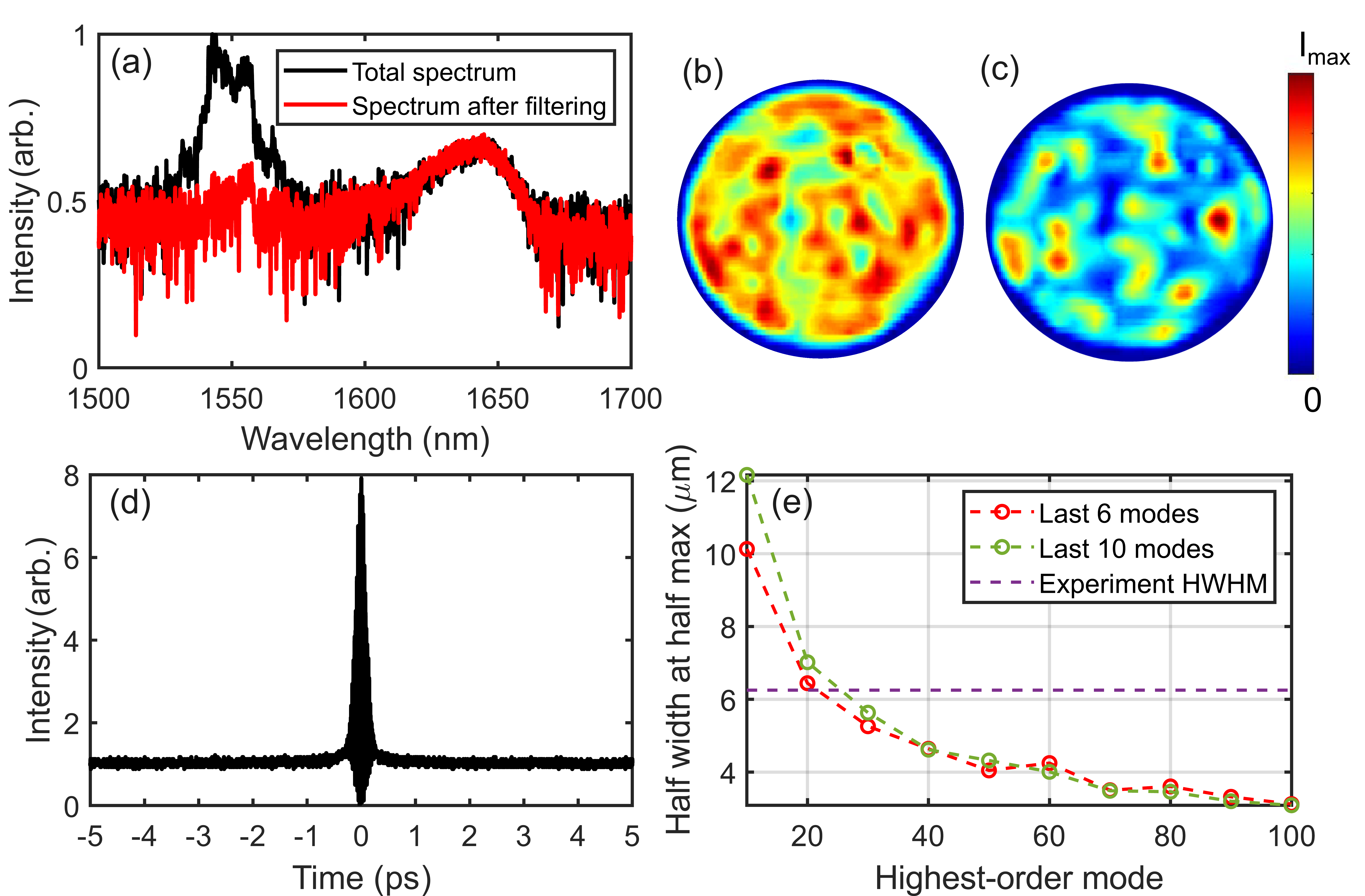}
\caption{Isolation of MM soliton with a long-pass filter. (a) Spatially-integrated spectrum before (black curve) and after (red curve) the filter on the normalized logarithmic intensity scale. (b) Spatial profile of the output before filtering. (c) Spatial profile of the output after filtering. (d) Spatially-integrated pulse measurement. (e) Simulated half-width at half-maximum (HWHM) of the spatial correlation function for randomly-generated beam patterns that arise from the last 6 modes (red) and from the last 10 modes (green) with the highest-order mode shown in the x-axis.  The purple dashed line is the HWHM of the experimental spatial correlation function.}
\label{Figure_5_Soliton Isolation & Mode Estimation}
\end{figure}

MM solitons can be isolated by filtering out the $\sim 1550 $ nm spectral peak with a long-pass filter with a cut-off at 1600 nm. A representative example is shown in Fig.~\ref{Figure_5_Soliton Isolation & Mode Estimation}, where the input pulse energy is 320 nJ and the Raman peak is centered at 1650 nm. Before filtering (Fig.~\ref{Figure_5_Soliton Isolation & Mode Estimation}(b)), the output intensity has low speckle contrast ($C = 0.17$) due to the temporal integration over multiple incoherent waves by the camera. After filtering away the dispersive waves (Fig.~\ref{Figure_5_Soliton Isolation & Mode Estimation}(c)), a marked increase in contrast is observed ($C = 0.42$), which is consistent with the numerical results for a soliton. The pulse after filtering has 44 nJ energy, a pulse width of 120 fs, and no discernable secondary temporal structures (Fig.~\ref{Figure_5_Soliton Isolation & Mode Estimation}(d)) (see supplementary materials for a measurement with a larger temporal range). The peak power of the MM soliton is $\sim 370$ kW. We choose this example because the presence of the second Raman peak at higher input pulse energies makes it difficult to ascertain the peak power of each soliton.

We approximately estimate the modal contents of observed solitons by combining information from spatial intensity correlations with the spectro-temporal properties of the pulses.  As indicated above, numerical simulations show that solitons tend to form in a group of modes with adjacent indices, as one might expect intuitively considering the need to overcome the temporal separation of modes in linear propagation.  The time-bandwidth product (TBP) can be used as a measure of the range of occupied modes because a wider group of modes will require greater spectral shifts between the modes to compensate for the increase in modal delay. With approximately constant pulse duration in each mode, this will yield a larger TBP. The soliton of Fig.~\ref{Figure_5_Soliton Isolation & Mode Estimation} is a representative example. The time-bandwidth product for the filtered pulse is 0.39. (For reference, the TBP of a transform-limited sech-shaped pulse is 0.315.) In simulations, a 6-mode soliton has a TBP of 0.36, and a 10-mode soliton has a TBP of 0.46 (see supplementary materials). The experimental value (0.39) is between the theoretical values for 6- and 10-mode solitons. To determine the mode indices of the band, we use the spatial intensity correlation function. In step-index fiber, nondegenerate modes have different spatial frequencies: the higher the mode index, the higher the spatial frequency. As a consequence, the length scale of the spatial correlation function of the intensity pattern will be related to the mode content; a narrower spatial correlation peak corresponds to higher-order mode content (see supplementary materials). The half-width at half-maximum (HWHM) of the spatial correlation function of the measured MM soliton in Fig. \ref{Figure_5_Soliton Isolation & Mode Estimation}(c) is around 6.2 \textmu m. We compare this to the HWHM of simulated patterns (see supplementary materials for details) with 6 modes and 10 modes in Fig. \ref{Figure_5_Soliton Isolation & Mode Estimation}(e). Both approach the experimental value when the index of the highest mode is in the range 22-25. Thus, we estimate that the MM soliton in the experiment contains 6-10 modes, with the highest order mode around 22-25. This estimate is consistent with the results of numerical simulations in Fig.~\ref{Figure_2_Modal_TS}. Direct measurements of the modal contents of the solitons will be a valuable future direction for this work.

Numerical simulations predict the main experimental features including the speckled spatial profile, the spectra, and the soliton duration. However, there are some discrepancies between simulated and experimental results. The energy fraction in the Raman solitons is about 14\% in the experiments, which is significantly less than in the simulations ($\sim 50$\%). Another observation is that the speckle contrast of MM solitons in the experiments is lower than in simulations. Bending-induced linear mode-coupling may underlie these discrepancies ~\cite{Flaes2018}. Even the loose coiling employed in the experiments may introduce mode-coupling that is neglected in the simulations. A preliminary experiment was performed with stronger mode-coupling, by coiling the fiber on a smaller (15-cm diameter) spool. Although similar spectra are observed, the intensity profiles after the long-pass filter have lower contrast than those obtained with loose coiling (see supplementary materials). Linear mode-coupling would be expected to eventually transfer energy to modes that are not part of the soliton. Due to the delay between modes, the camera would record an incoherent sum of the mode intensities, with reduced contrast. More systematic studies will be needed to understand the effects of mode-coupling on MM solitons in step-index fiber.

The MM solitons observed in step-index fiber have 3 times higher peak power than solitons observed to date in GRIN fiber \cite{Wright2015}. There may be applications of the high-power, femtosecond pulses with speckled intensity profiles that are produced by MM soliton formation. MM solitons in step-index fiber can reach high power thanks to the large effective mode areas along with the high power required for cross-phase modulation to compensate for relatively-large modal dispersion. It will be interesting to study the limits of high-power and highly-multimode solitons in the future.  

To summarize, highly-multimode solitons are observed in step-index fiber. Numerical simulations and experimental results are consistent with the generation of solitons that consist of superpositions of 6 to 10 high-order transverse modes. The resulting pulses have modal occupancies that are temporally aligned, yet can be spatio-spectrally complex. Their duration is around 100 fs and the peak power can exceed 300 kW. Knowledge of solitons in step-index fiber will help provide a framework for a variety of multimode nonlinear phenomena and may be relevant to space-division multiplexing communication systems and applications in imaging. 

\bibliographystyle{unsrt}
\bibliography{main}

\begin{thebibliography}{10}

\bibitem{Agrawal2000}
Govind~P Agrawal.
\newblock {\em Nonlinear fiber optics}.
\newblock Elsevier, 2013.

\bibitem{Hasegawa1973}
Akira Hasegawa and Frederick Tappert.
\newblock Transmission of stationary nonlinear optical pulses in dispersive
  dielectric fibers. i. anomalous dispersion.
\newblock {\em Applied Physics Letters}, 23(3):142--144, 1973.

\bibitem{Mollenauer1980}
Linn~F Mollenauer, Roger~H Stolen, and James~P Gordon.
\newblock Experimental observation of picosecond pulse narrowing and solitons
  in optical fibers.
\newblock {\em Physical Review Letters}, 45(13):1095, 1980.

\bibitem{Herrmann2002}
J~Herrmann, U~Griebner, N~Zhavoronkov, A~Husakou, D~Nickel, JC~Knight,
  WJ~Wadsworth, P~St~J Russell, and G~Korn.
\newblock Experimental evidence for supercontinuum generation by fission of
  higher-order solitons in photonic fibers.
\newblock {\em Physical review letters}, 88(17):173901, 2002.

\bibitem{Skryabin2010}
Dmitry~V Skryabin and Andrey~V Gorbach.
\newblock Colloquium: Looking at a soliton through the prism of optical
  supercontinuum.
\newblock {\em Reviews of Modern Physics}, 82(2):1287, 2010.

\bibitem{Mollenauer1984}
Linn~F Mollenauer and Roger~H Stolen.
\newblock The soliton laser.
\newblock {\em Optics letters}, 9(1):13--15, 1984.

\bibitem{Haus1985}
H~Haus and M~Islam.
\newblock Theory of the soliton laser.
\newblock {\em IEEE journal of quantum electronics}, 21(8):1172--1188, 1985.

\bibitem{Hasegawa2000}
Akira Hasegawa.
\newblock Soliton-based optical communications: An overview.
\newblock {\em IEEE Journal of Selected Topics in Quantum Electronics},
  6(6):1161--1172, 2000.

\bibitem{Mafi2012}
Arash Mafi.
\newblock Pulse propagation in a short nonlinear graded-index multimode optical
  fiber.
\newblock {\em Journal of Lightwave Technology}, 30(17):2803--2811, 2012.

\bibitem{Hasegawa1980}
Akira Hasegawa.
\newblock Self-confinement of multimode optical pulse in a glass fiber.
\newblock {\em Optics letters}, 5(10):416--417, 1980.

\bibitem{Crosignani1981}
Bruno Crosignani and Paolo Di~Porto.
\newblock Soliton propagation in multimode optical fibers.
\newblock {\em Optics letters}, 6(7):329--330, 1981.

\bibitem{Yu1995}
Shinn-Sheng Yu, Chih-Hung Chien, Yinchieh Lai, and Jyhpyng Wang.
\newblock Spatio-temporal solitary pulses in graded-index materials with kerr
  nonlinearity.
\newblock {\em Optics communications}, 119(1-2):167--170, 1995.

\bibitem{Chien1996}
Chih-Hung Chien, Shinn-Sheng Yu, Yinchieh Lai, and Jyhpyng Wang.
\newblock Off-axial pulse propagation in graded-index materials with kerr
  nonlinearity—a variational approach.
\newblock {\em Optics communications}, 128(1-3):145--157, 1996.

\bibitem{Raghavan2000}
S~Raghavan and Govind~P Agrawal.
\newblock Spatiotemporal solitons in inhomogeneous nonlinear media.
\newblock {\em Optics Communications}, 180(4-6):377--382, 2000.

\bibitem{Lederer2008}
Falk Lederer, George~I Stegeman, Demetri~N Christodoulides, Gaetano Assanto,
  Moti Segev, and Yaron Silberberg.
\newblock Discrete solitons in optics.
\newblock {\em Physics Reports}, 463(1-3):1--126, 2008.

\bibitem{Grudinin1988}
AB~Grudinin, EM~Dianov, DV~Korbkin, AM~Prokhorov, and DV~Khaǐdarov.
\newblock Nonlinear mode coupling in multimode optical fibers; excitation of
  femtosecond-range stimulated-raman-scattering solitons.
\newblock {\em Soviet Journal of Experimental and Theoretical Physics Letters},
  47:356, 1988.

\bibitem{Zitelli2021}
Mario Zitelli, Fabio Mangini, Mario Ferraro, Oleg Sidelnikov, and Stefan
  Wabnitz.
\newblock Conditions for walk-off soliton generation in a multimode fiber.
\newblock {\em Communications Physics}, 4(1):182, 2021.

\bibitem{Wright2015}
Logan~G Wright, William~H Renninger, Demetrios~N Christodoulides, and Frank~W
  Wise.
\newblock Spatiotemporal dynamics of multimode optical solitons.
\newblock {\em Optics express}, 23(3):3492--3506, 2015.

\bibitem{Sun2022}
Yifan Sun, Mario Zitelli, Mario Ferraro, Fabio Mangini, Pedro Parra-Rivas, and
  Stefan Wabnitz.
\newblock Multimode soliton collisions in graded-index optical fibers.
\newblock {\em Optics Express}, 30(12):21710--21724, 2022.

\bibitem{Renninger2013}
William~H Renninger and Frank~W Wise.
\newblock Optical solitons in graded-index multimode fibres.
\newblock {\em Nature communications}, 4(1):1719, 2013.

\bibitem{rishoj2019}
L~Rish{\o}j, Boyin Tai, Poul Kristensen, and Siddharth Ramachandran.
\newblock Soliton self-mode conversion: revisiting raman scattering of
  ultrashort pulses.
\newblock {\em Optica}, 6(3):304--308, 2019.

\bibitem{Zitelli2022}
Mario Zitelli, Yifan Sun, Mario Ferraro, Fabio Mangini, Oleg Sidelnikov,
  Vincent Couderc, and Stefan Wabnitz.
\newblock Multimode solitons in step-index fibers.
\newblock {\em Optics Express}, 30(4):6300--6310, 2022.

\bibitem{Terry2007}
Nathan~B Terry, Thomas~G Alley, and Timothy~H Russell.
\newblock An explanation of srs beam cleanup in graded-index fibers and the
  absence of srs beam cleanup in step-index fibers.
\newblock {\em Optics express}, 15(26):17509--17519, 2007.

\bibitem{Wu2022}
Yuhang Wu, Demetrios~N Christodoulides, and Frank~W Wise.
\newblock Multimode nonlinear dynamics in spatiotemporal mode-locked
  anomalous-dispersion lasers.
\newblock {\em Optics Letters}, 47(17):4439--4442, 2022.

\bibitem{Poletti2008}
Francesco Poletti and Peter Horak.
\newblock Description of ultrashort pulse propagation in multimode optical
  fibers.
\newblock {\em JOSA B}, 25(10):1645--1654, 2008.

\bibitem{Wright2017}
Logan~G Wright, Zachary~M Ziegler, Pavel~M Lushnikov, Zimu Zhu, M~Amin
  Eftekhar, Demetrios~N Christodoulides, and Frank~W Wise.
\newblock Multimode nonlinear fiber optics: massively parallel numerical
  solver, tutorial, and outlook.
\newblock {\em IEEE Journal of Selected Topics in Quantum Electronics},
  24(3):1--16, 2017.

\bibitem{Bromberg2014}
Yaron Bromberg and Hui Cao.
\newblock Generating non-rayleigh speckles with tailored intensity statistics.
\newblock {\em Physical Review Letters}, 112(21):213904, 2014.

\bibitem{Flaes2018}
Dirk E~Boonzajer Flaes, Jan Stopka, Sergey Turtaev, Johannes~F De~Boer,
  Tom{\'a}{\v{s}} Tyc, and Tom{\'a}{\v{s}} {\v{C}}i{\v{z}}m{\'a}r.
\newblock Robustness of light-transport processes to bending deformations in
  graded-index multimode waveguides.
\newblock {\em Physical review letters}, 120(23):233901, 2018.

\end{thebibliography}

\end{document}


\date{}
\maketitle

\section{Modal delay in step-index multimode fiber}

\begin{figure}[ht]
\centering
\includegraphics[width=0.8\linewidth]{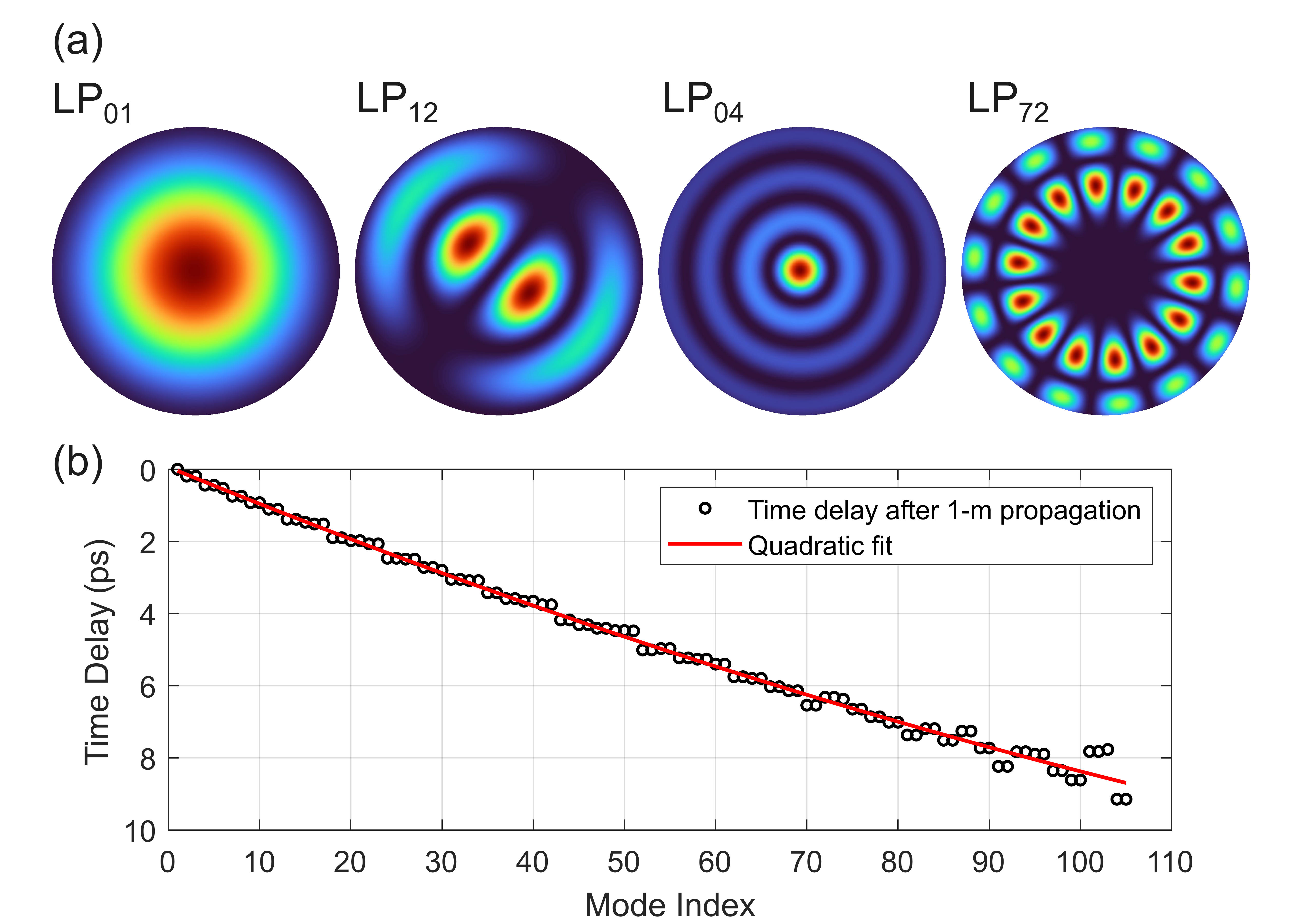}
\caption{Modal delay in SI MMF. (a) Intensity profiles of representative modes. (b) Time delay for different modes under linear pulse propagation. The red curve is a quadratic fit to the data points.}
\label{SUPPFigure_1_Step-Index Fiber}
\end{figure}

Here, we want to show the modal delay in step-index (SI) MM fiber. Fig. \ref{SUPPFigure_1_Step-Index Fiber}(a) shows several example eigenmodes of SI MM fiber. Each mode has a different propagation constant, group velocity, and group velocity dispersion. In Fig. \ref{SUPPFigure_1_Step-Index Fiber}(b), the modal delay with respect to the fundamental mode is plotted for 1 m propagation in the fiber used in the experiments, which supports 105 modes in each polarization. The largest time delay is for mode 105, $\sim 9$ ps. The time delay increases monotonically with the increase of the mode index. We apply a quadratic fit to the points and obtain the red curve in Fig. \ref{SUPPFigure_1_Step-Index Fiber}(b):

\begin{equation}
    T = -1.937 \times 10^{-4} M^2 + 0.103 M -0.0603
\end{equation}

with T the time delay and M the mode index. The linear term is dominant and has an average slope of 103 fs/m. 

\newpage

\section{Linear pulse propagation }

\begin{figure}[ht]
\centering
\includegraphics[width=\linewidth]{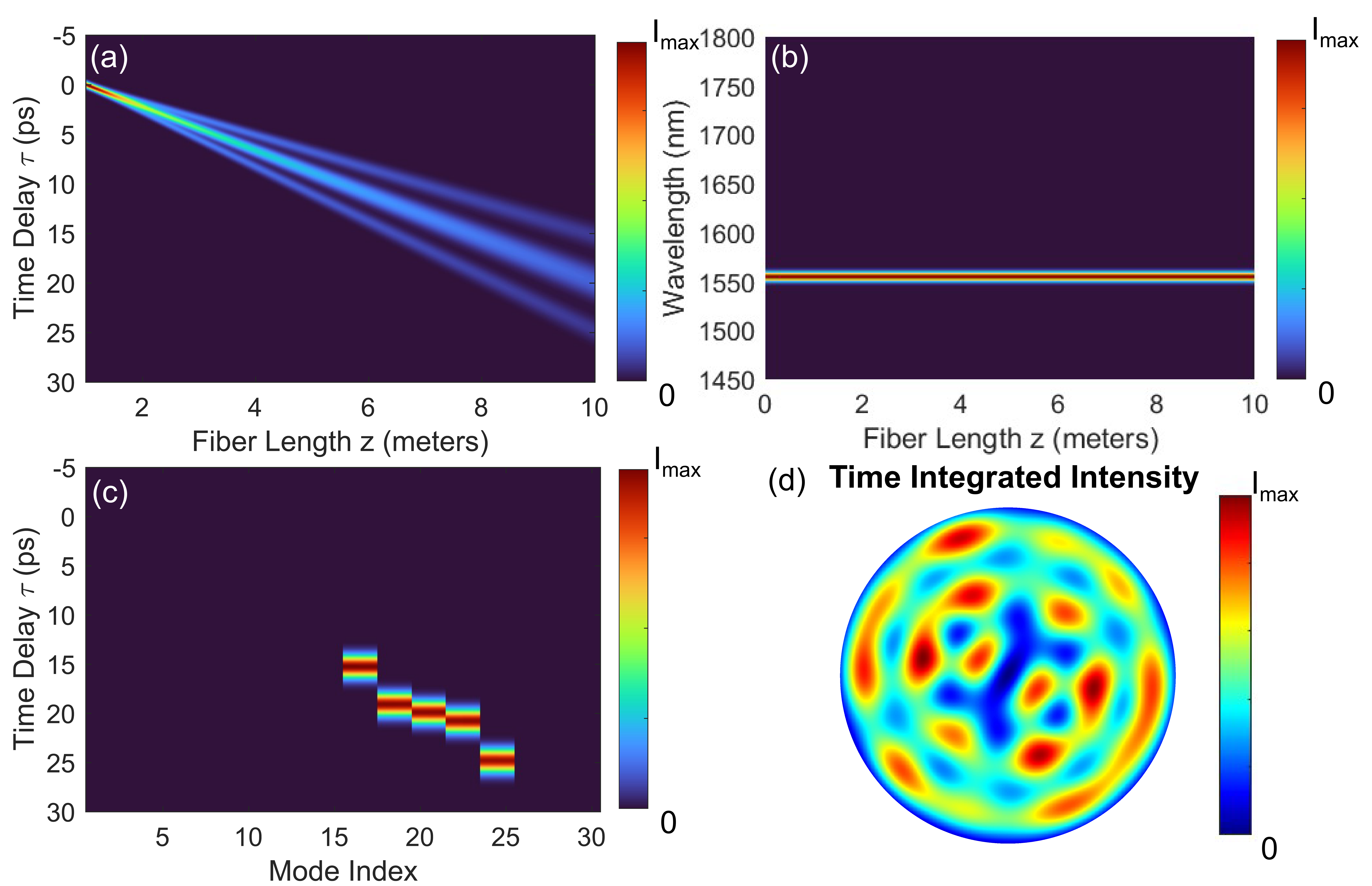}
\caption{Linear pulse propagation in a 10-m SI MMF. (a) Spatially-averaged temporal profile versus propagation distance. (b) Spatially-averaged spectral profile versus propagation distance. (c) Temporal profile in each mode at the end of the fiber. (d) Time-integrated intensity beam profile.}
\label{SUPPFigure_2_Linear Pulse Propagation}
\end{figure}

Pulse propagation with negligible nonlinearity is shown in Fig. \ref{SUPPFigure_2_Linear Pulse Propagation}. The launched mode contents are the same as what is simulated in the nonlinear case in the main text. The total input pulse energy is 0.1 nJ. The pulses in different modes will broaden under group velocity dispersion (Fig. \ref{SUPPFigure_2_Linear Pulse Propagation}(a)) and will separate because of the modal dispersion (Fig. \ref{SUPPFigure_2_Linear Pulse Propagation}(c)), while the spectrum will remain unchanged (Fig. \ref{SUPPFigure_2_Linear Pulse Propagation}(b)). The beam profile has the contrast $C=0.39$.

\newpage

\section{Examples of MM Solitons in Simulations}

The MM solitons shown in the main text occupy higher-order modes, with negligible power in the lower-order modes.  In simulations it is also possible to observe MM solitons that occupy lower-order modes in simulations. A couple of examples are shown below. 

Fig. \ref{SUPPFigure_3_Second SI MMS} shows a soliton that occupies primarily the lowest 6 modes of the fiber. It has a pulse width of $\sim 100$ fs and a time-bandwidth product of 0.36, with the spectrum centered at $\sim 1650$ nm.  

\begin{figure}[ht]
\centering
\includegraphics[width=\linewidth]{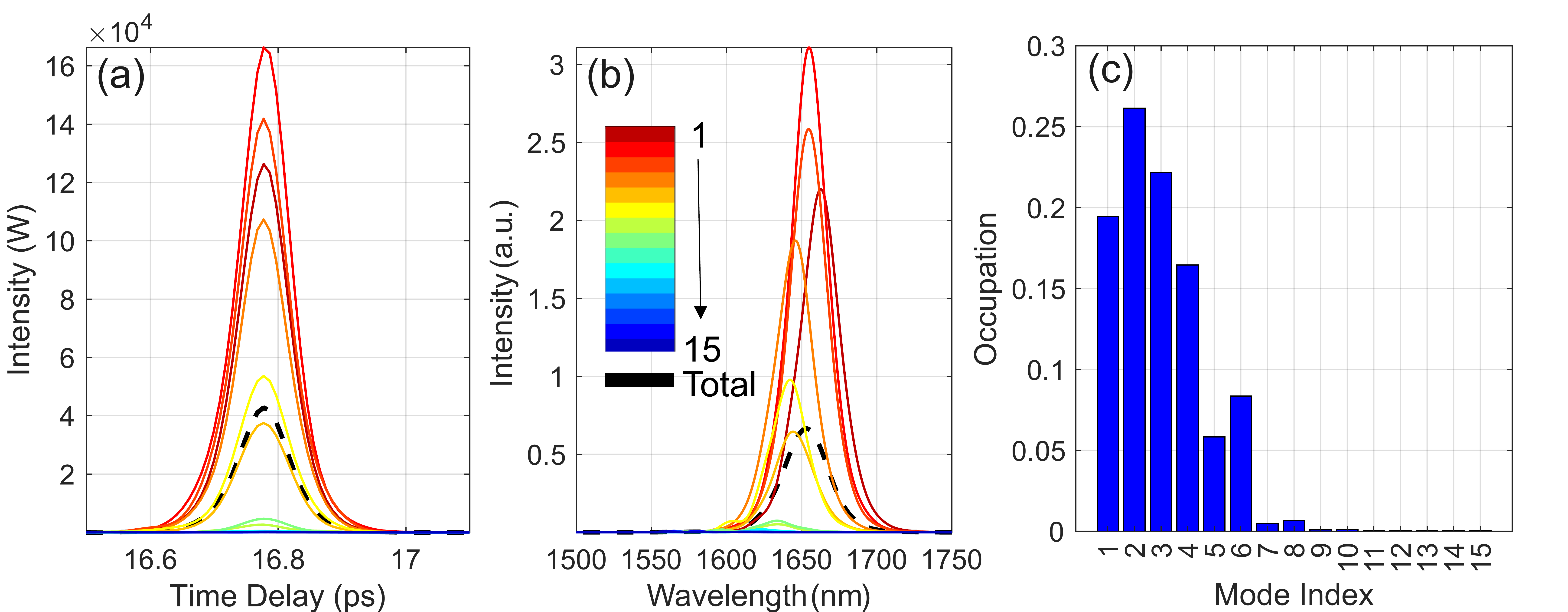}
\caption{Mode-resolved temporal and spectral shapes for a 6-mode soliton. In (a), the dashed black curve is the total temporal profile normalized by a factor of 15. The colored curves are mode-resolved temporal profiles. In (b), the dashed black curve is the total spectral profile normalized by a factor of 15. The colored curves are mode-resolved spectral profiles. (c) is the modal occupation.}
\label{SUPPFigure_3_Second SI MMS}
\end{figure}

\begin{figure}[ht]
\centering
\includegraphics[width=\linewidth]{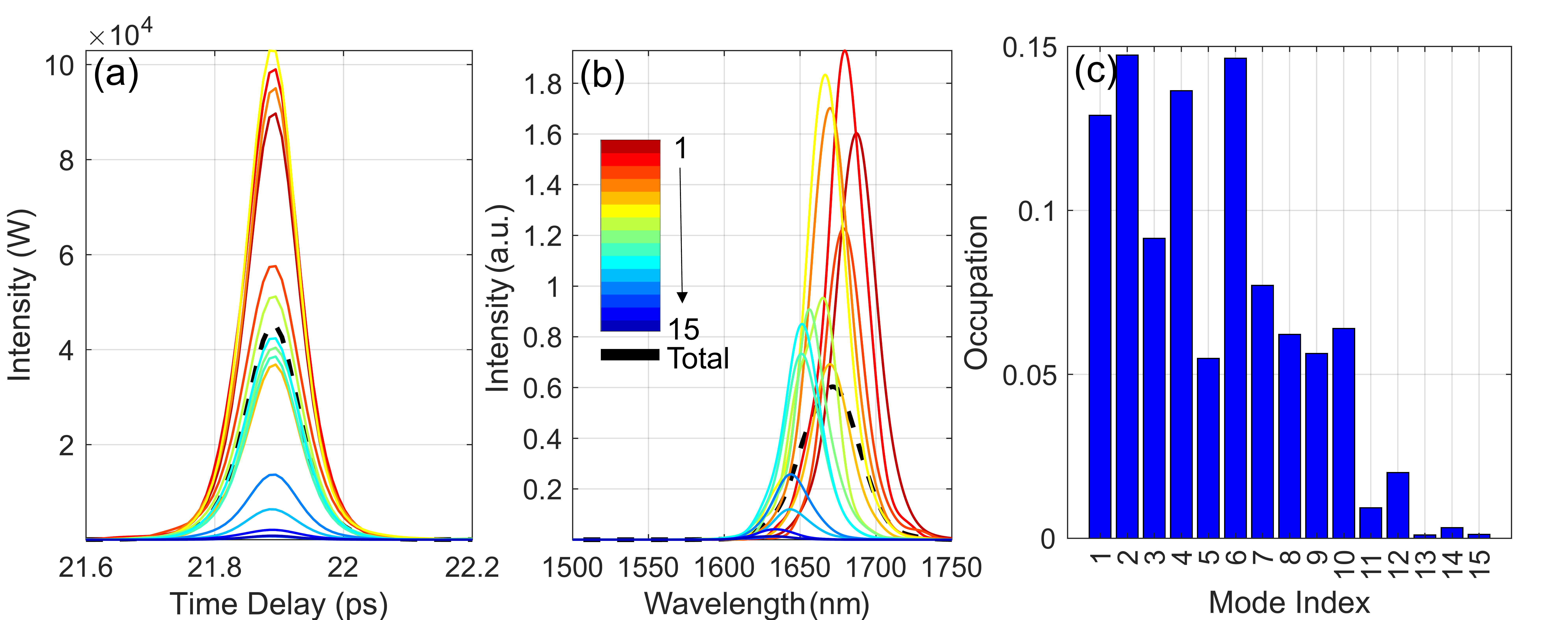}
\caption{Mode-resolved temporal and spectral shapes for a 10-mode soliton. In (a), the dashed black curve is the total temporal profile normalized by a factor of 15. The colored curves are mode-resolved temporal profiles. In (b), the dashed black curve is the total spectral profile normalized by a factor of 15. The colored curves are mode-resolved spectral profiles. (c) is the modal occupation.}
\label{SUPPFigure_4_Modal_TS}
\end{figure}

Fig. \ref{SUPPFigure_4_Modal_TS} shows a soliton that occupies primarily the lowest 10 modes of the fiber. It has a pulse width of $\sim 100$ fs and a time-bandwidth product of 0.36, with the spectrum centered at $\sim 1660$ nm.   
 
The time-bandwidth products will be used for estimating the modal occupancy of experimental solitons.

\newpage

\section{Long-range autocorrelation measurements and spatial beam profiles}

\begin{figure}[ht]
\centering
\includegraphics[width=1\linewidth]{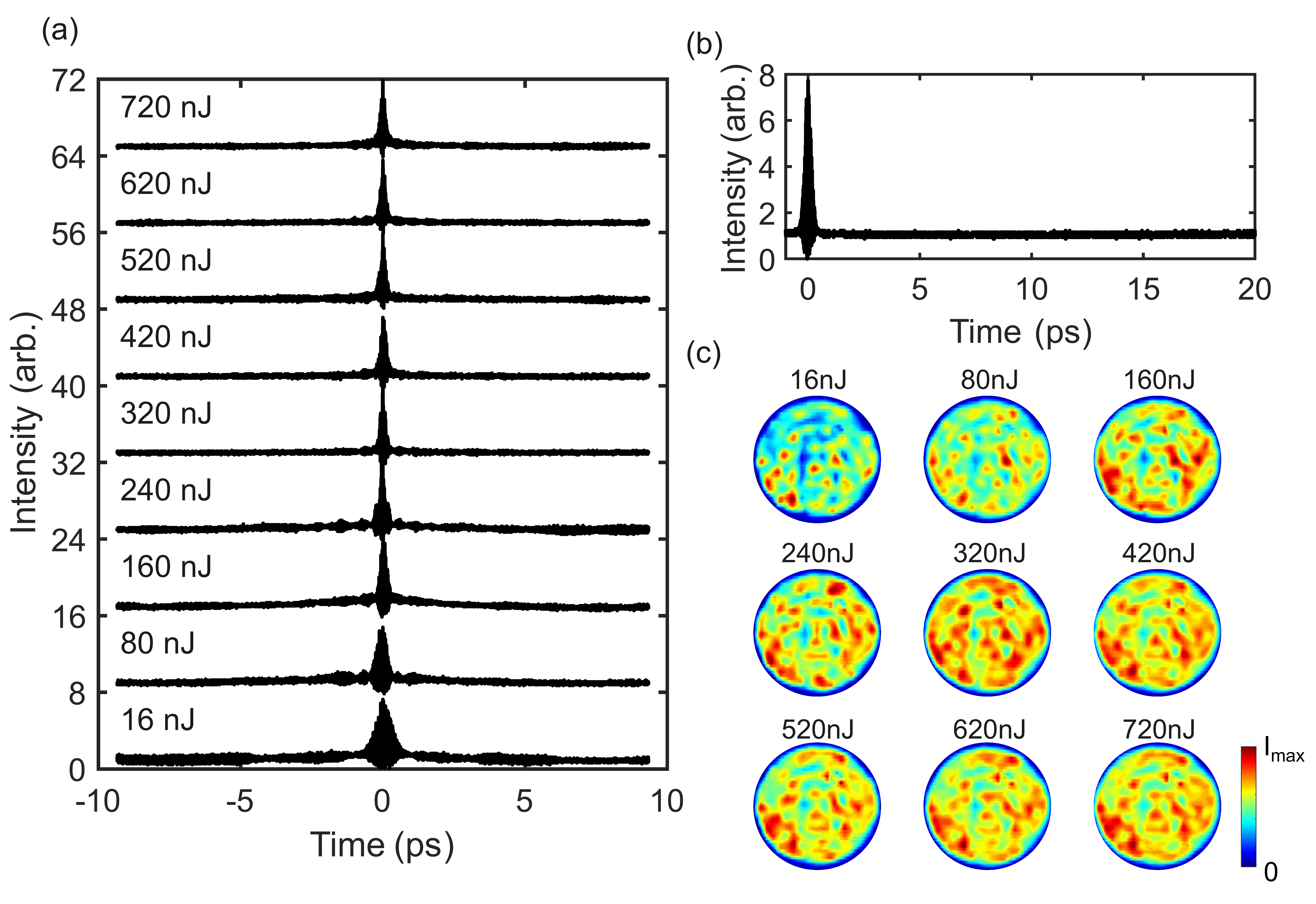}
\caption{Long-range pulse measurement and temporal-integrated near-field spatial profiles. (a) Two-photon induced trace for output beam at different input pulse energy. (b) Long-range autocorrelation measurement for the case with 320 nJ input pulse energy and a long-pass filter at the output. (c) The time-integrated spatial beam profiles at different input pulse energies.}
\label{SUPPFigure_5_Spatial Profiles}
\end{figure}

The results for long-range beam autocorrelation at different input pulse energies are shown in Fig. \ref{SUPPFigure_5_Spatial Profiles} (a). The measuring range is as large as 10 ps. There are no additional structures in the 5 - 10 ps range. Fig. \ref{SUPPFigure_5_Spatial Profiles} (b) shows the long-range autocorrelation for the soliton isolated and discussed in the main text. The measuring range is as large as 20 ps. There are no additional structures in the 5 - 20 ps range.

The spatial beam profiles corresponding to the pulse and spectrum measurement in the main text are shown in Fig. \ref{SUPPFigure_5_Spatial Profiles} (c). The beam profiles up to 720 nJ input pulse energy are speckled, with low contrast ratio. 

\newpage

\section{Calculation of spatial correlation functions for speckled patterns}

\begin{figure}[ht]
\centering
\includegraphics[width=0.95\linewidth]{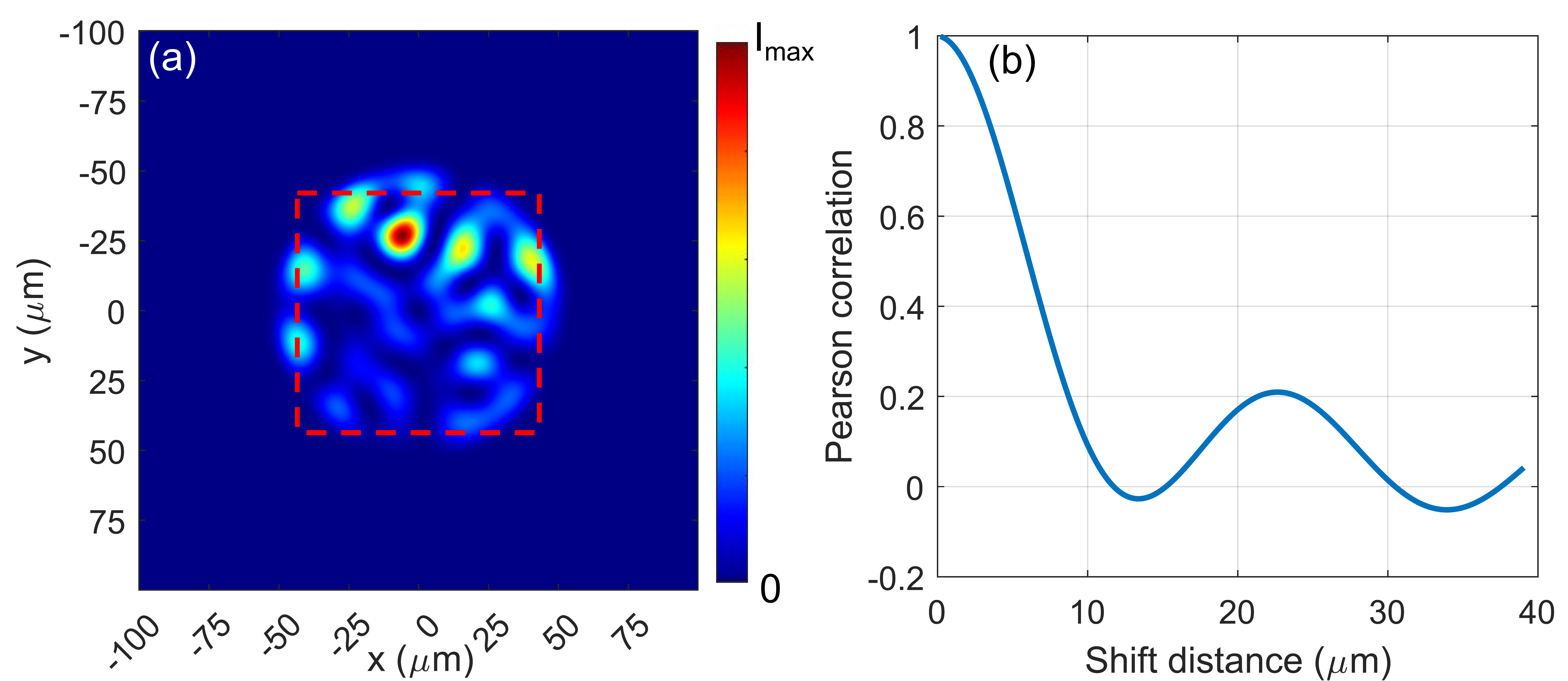}
\caption{Spatial correlation function for a multimode beam pattern. (a) A simulated multimode beam pattern. The red dashed square is the region used to calculate the spatial correlation. (b) The spatial correlation function for this multimode beam pattern.}
\label{SUPPFigure_6_Spatial Correlation function}
\end{figure}

We want to give a detailed demonstration of spatial correlation functions used to estimate modal occupancy. In Fig. \ref{SUPPFigure_6_Spatial Correlation function}(a), we show an example multimode beam pattern. It has equal amplitudes in modes 16-25, but random phases. The red dashed square selects the region to calculate the spatial correlation. We calculate the Pearson Correlation,

\begin{equation}
    \rho_{X,Y} =  \frac{\EX[(X-\mu_X)(Y-\mu_Y)]}{\sigma_X \sigma_Y}
\end{equation}

In this equation, $\EX[X_0]$ means the expectation value of the variable $X_0$, $X$ is the array of the original intensity pattern, and $Y$ is the array of the intensity pattern that shifts a particular distance along the x direction with respect to the original one. (Patterns that shift outside the red square at one edge would shift back from the opposite edge.) $\mu$ and $\sigma$ correspond to the average value and standard deviation of $X$ or $Y$. The correlation value with respect to the shift distance is shown in Fig. \ref{SUPPFigure_6_Spatial Correlation function}(b). The half-width-at-half-maximum (HWHM) is defined as the shift distance corresponding to the 0.5 correlation value. In Fig. 5(e), each data point is the averaged HWHM of 1000 beam patterns. They have the same amplitude distribution in each mode but random phase relations. Similar results are obtained for shifts in any direction.

\newpage

\section{Experiment results for small fiber coiling radius}

\begin{figure}[ht]
\centering
\includegraphics[width=0.8\linewidth]{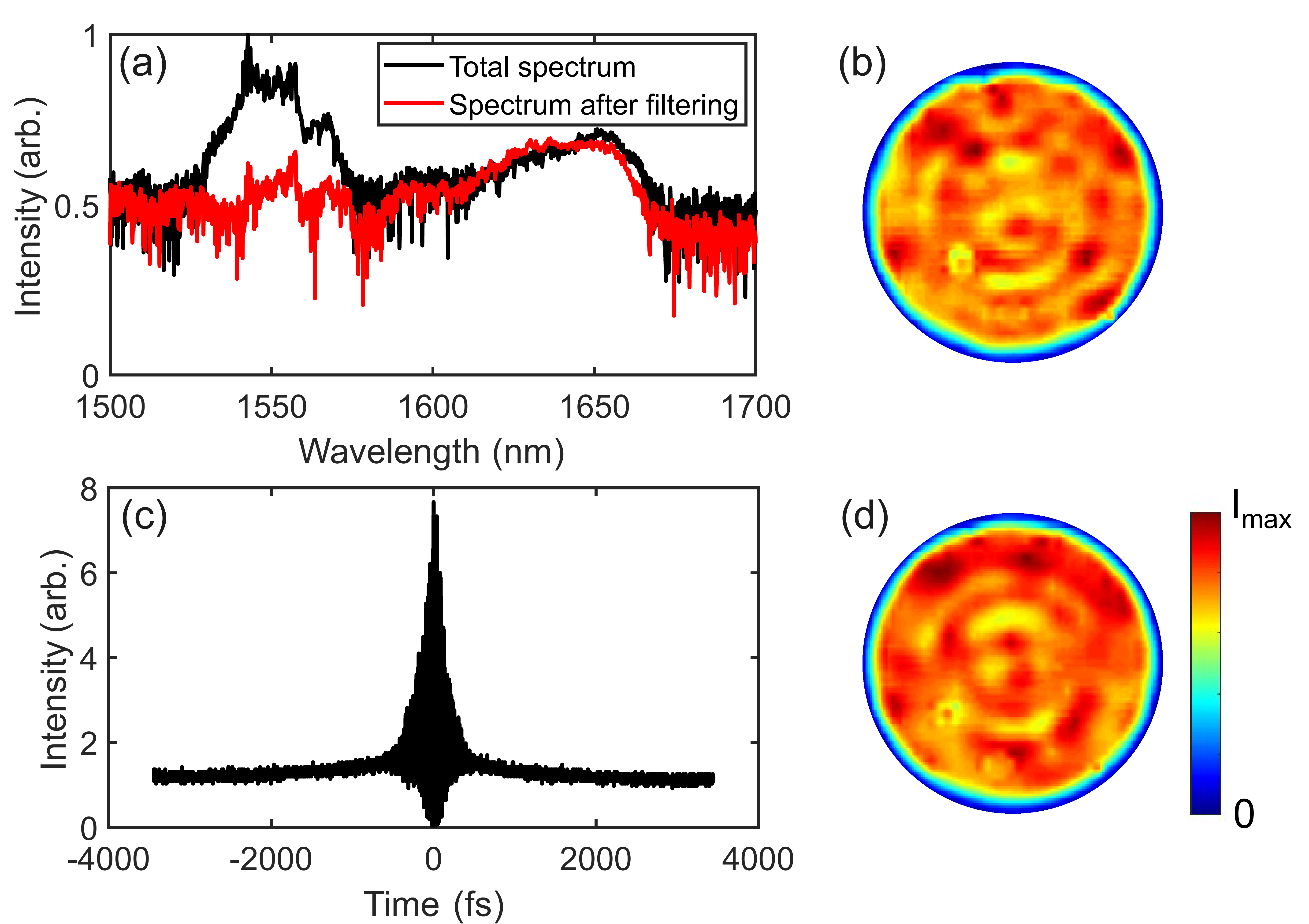}
\caption{Measurement results for 10 m of fiber coiled on a 15-cm diameter spool. (a) The spatially-averaged spectrum before and after the long-pass filter with 1600 nm edge. (b) Spatial profile for the total spectrum. (c) Two-photon autocorrelation trace for the filtered spectrum. (d) Spatial profile after the long-pass filter.}
\label{SUPPFigure_7_Smaller Fiber Coil}
\end{figure}

In Fig. \ref{SUPPFigure_7_Smaller Fiber Coil}, we show the measurement results with 330-nJ and 500-fs pulses launched into the fiber, which is coiled with a diameter of 15 cm. The spectrum (Fig. \ref{SUPPFigure_7_Smaller Fiber Coil}(a)) has a distinguishable Raman peak at 1650 nm. After the long pass filter, the Raman peak is dominant. The pulse width after filtering is 160 fs (Fig. \ref{SUPPFigure_7_Smaller Fiber Coil}(c)). Fig. \ref{SUPPFigure_7_Smaller Fiber Coil}(b) and S8(d) are the beam profiles before and after filtering. The beam after filtering consistently exhibits lower contrast than the one shown in Fig. 5. As discussed in the main text, a possible explanation is that tighter coiling leads to a higher linear mode coupling. Thus, the soliton energy constantly couples to the other modes which finally become dispersive waves, and the mode patterns are added incoherently.

\newpage

\section{Generation of MM solitons with smaller speckle grains}

\begin{figure}[ht]
\centering
\includegraphics[width=0.8\linewidth]{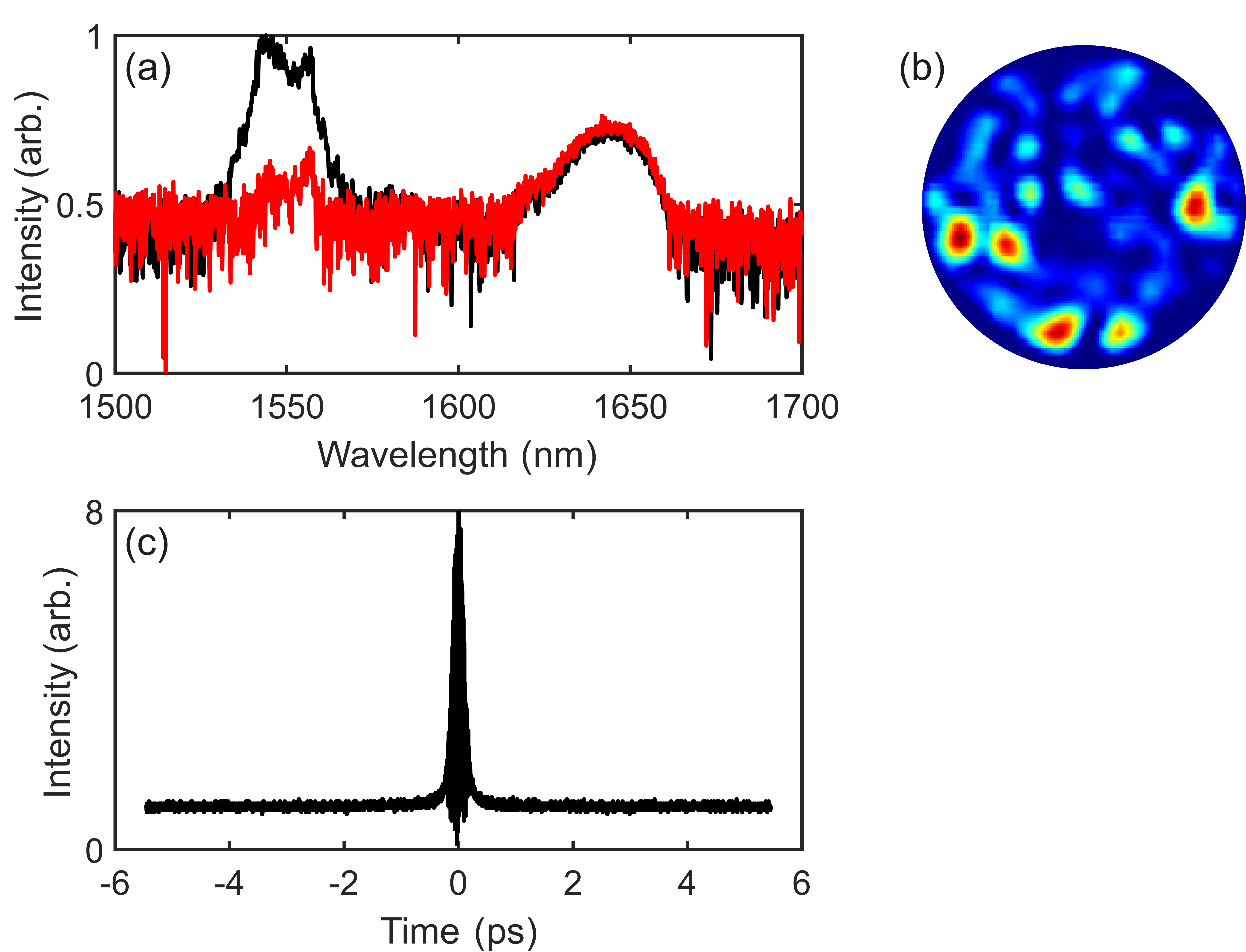}
\caption{Measurement results for MM soliton with smaller speckled grain size. (a) Spatially-integrated spectrum before (black) and after (red) the long-pass filter. (b) Spatial intensity profile. (c) Spatially-integrated autocorrelation trace.}
\label{SUPPFigure_8_Another Example}
\end{figure}

MM solitons with smaller speckle grains than shown in the main text can be generated, which also illustrates that MM soliton formation can occur in other groups of modes.  We change the launching condition by deviating the input beam further from the center of MMF. The input pulse energy is 320 nJ. The measured spectra before and after filtering with a long-pass filter are shown in Fig. \ref{SUPPFigure_8_Another Example}(a). The spatial beam profile with a speckled pattern is shown in Fig. \ref{SUPPFigure_8_Another Example}(b). The HWHM of the spatial correlation function is 5.5 \textmu m, which from fig. 5 in the main text corresponds to a superposition of modes 30-33. The pulse duration is measured to be 130 fs and the autocorrelation trace is shown in Fig. \ref{SUPPFigure_8_Another Example} (c).

